\documentclass{article}

\usepackage{arxiv}

\usepackage[utf8]{inputenc} % allow utf-8 input
\usepackage[T1]{fontenc}    % use 8-bit T1 fonts
\usepackage{hyperref}       % hyperlinks
\usepackage{url}            % simple URL typesetting
\usepackage{booktabs}       % professional-quality tables
\usepackage{amsfonts}       % blackboard math symbols
\usepackage{nicefrac}       % compact symbols for 1/2, etc.
\usepackage{microtype}      % microtypography
\usepackage{lipsum}		% Can be removed after putting your text content
\usepackage{graphicx}
\usepackage{amsmath}
\usepackage{esvect}
\usepackage{doi}

\title{Intrinsic mono-chromatic emission of X and gamma-rays in symmetric electron-photon beam collisions}

	\author{\href{https://orcid.org/0000-0002-4367-1555}{\includegraphics[scale=0.06]{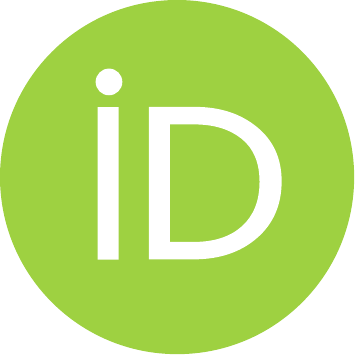}\hspace{1mm}L. Serafini} \\
	INFN-Section of Milan\\
	Via G. Celoria 16\\
	Milano, Italy, 20133 \\
    \And
	\href{https://orcid.org/0000-0002-8556-3384}{\includegraphics[scale=0.06]{orcid.pdf}\hspace{1mm}V. Petrillo} \\
    INFN-Section of Milan\\
	Via G. Celoria 16\\
	Milano, Italy, 20133 \\
    \And    
    \href{https://orcid.org/0000-0001-6010-9225}{\includegraphics[scale=0.06]{orcid.pdf}\hspace{1mm}A. Bacci} \\
    INFN-Section of Milan\\
	Via G. Celoria 16\\
	Milano, Italy, 20133 \\
    \And
    \href{https://orcid.org/0000-0003-0876-3830}{\includegraphics[scale=0.06]{orcid.pdf}\hspace{1mm}C. Curatolo} \\
    INFN-Section of Milan\\
	Via G. Celoria 16\\
	Milano, Italy, 20133 \\
    \And 
     \href{https://orcid.org/0000-0002-9152-0102}{\includegraphics[scale=0.06]{orcid.pdf}\hspace{1mm}I. Drebot} \\
    INFN-Section of Milan\\
	Via G. Celoria 16\\
	Milano, Italy, 20133 \\
    \And
 \href{https://orcid.org/0000-0002-5767-3850}{\includegraphics[scale=0.06]{orcid.pdf}\hspace{1mm}M. Rossetti~Conti} \\
	INFN-Section of Milan\\
	Via G. Celoria 16\\
	Milano, Italy, 20133 \\
	\And
    \href{https://orcid.org/0000-0002-6216-8664}{\includegraphics[scale=0.06]{orcid.pdf}\hspace{1mm}A. R. Rossi} \\
    INFN-Section of Milan\\
	Via G. Celoria 16\\
	Milano, Italy, 20133 \\
}

\renewcommand{\shorttitle}{\textit{arXiv} Template}

\hypersetup{
pdftitle={Symmetric Compton Scattering},
pdfauthor={M. Rossetti Conti},
}

\begin{document}
\maketitle

\begin{abstract}
    This paper explores the transition between Compton Scattering and Inverse Compton Scattering (ICS), which is characterized by an equal exchange of energy and momentum between the colliding particles (electrons and photons). This regime has been called Symmetric Compton Scattering (SCS) and has the unique property of cancelling the energy-angle correlation of scattered photons, and, when the electron recoil is large, transferring mono-chromaticity from one colliding beam to the other, resulting in back-scattered photon beams that are intrinsically monochromatic. The paper suggests that large-recoil SCS or quasi-SCS can be used to design compact intrinsic monochromatic gamma-ray sources based on compact linacs, thus avoiding the use of GeV-class electron beams together with powerful laser/optical systems as those typically required for ICS sources.

\end{abstract}

% \keywords{First keyword \and Second keyword \and More}

\section{Introduction}
\label{sec:intro}
%	The quantum-like interaction of an electron with a photon was first discovered by Arthur Compton in 1922 and interpreted as a collision between two point-like particles. 
 %
The spectral red-shift observed when a X-ray pulse interacts with a carbon target was observed by Arthur Compton in 1922 \cite{ArthurC} and interpreted as an effect of the collision between the photons of the X-rays and the electrons of the solid, both assumed as point-like particles.
The scattering of energetic photons by electrons at rest in the laboratory was called Compton effect after him.

Much later, the Inverse Compton Scattering (ICS) effect was studied \cite{KULIKOV1964344} and experimentally demonstrated at particle accelerators \cite{Ladon}, using highly relativistic electrons colliding with laser beams, within an inverse kinematics set-up where the electron looses energy and momentum in favor of the incident photon, that is back-scattered and up-shifted to much larger energies.

While the Compton effect cannot be explained classically, the low recoil regime of ICS, in which the electron energy/momentum loss is negligible, 
has been described in the framework of the classical electro-dynamics and it is known as Thomson effect \cite{Tomassini2008}.% 
%can be described easily and completely by classical electro-dynamics, following a formalism completely similar to that used for emission of Thomson radiation. 
In this paper we analyze the transition between Compton effect and ICS, occurring when the colliding particles exchange an equal amount of energy and momentum, and we call this regime Symmetric Compton Scattering (SCS). 
In this particular condition, the properties of the scattered photons are unique: unlike in all other radiations emitted with a Lorentz boost, SCS scattered photon energy indeed no longer depends on the scattering angle, so that the back-scattered radiation beam becomes intrinsically monochromatic. 
Extending the analyses on large electron recoil ICS carried out in Ref.  \cite{Recoil17} to this particular regime, we find that SCS is characterized by the transfer of mono-chromaticity from one colliding beam to the other, so that when a large bandwidth photon beam collides under SCS conditions with a mono-energetic electron beam, the back-scattered photon beam results to be mono-chromatized.
     
%A noticeable consequence of such an unique property of SCS (or quasi-SCS) at very large recoil is the possibility to  extend the photon energy range towards multi-MeV photons, accessing, in this way, the deep nuclear photonics/physics domain. (MARCELLO: ho commentato questa frase perché mi sembra collegata all'applicazione a valle di sorgenti FEL e Sincrotroni)
SCS or quasi-SCS at large recoil could allow to design compact sources of intrinsic mono-chromatic gamma-rays alimented by low energy MeV electron bunches, thus avoiding the use of GeV-class accelerators and powerful laser/optical systems, actually needed by ICS sources.

\section{Compton interaction regimes}
\label{sec:theory}
Considering the Compton interaction between photon pulses and counter-propagating electron beams, we can derive the well-known equation for the photon energy ($E_{\mathrm{ph}}^\prime=\hbar\omega^\prime$, with $\omega^\prime$ being the photon associated angular frequency and $\hbar$ the reduced Planck constant) scattered at an angle $\theta$. Following the notation of eq. 3 in Ref. \cite{Compton18}, we can write: 
\begin{equation}
\label{eq:GeneralOmega}
E_{\mathrm{ph}}^{\prime}(\theta)=\frac{(1+\beta) \gamma^2}{\gamma^2(1-\beta \cos \theta)+\frac{X}{4}(1+\cos \theta)}E_{\mathrm{ph}},
\end{equation}
where the incident photon energy is $E_{\mathrm{ph}}=\hbar\omega$, $\beta=v_e/c$ is the ratio between electron velocity $v_e$ and light speed $c$, $\gamma=1/\sqrt{1-\beta^2}$ is electron Lorentz factor and  $X$ is the electron recoil factor that introduces an important contribution at high energy of both incident photons and electrons. 
$X$ has been defined in \cite{Recoil17} (eq. 4) as: 
\begin{equation}
\label{eq:GeneralRecoil}
X=\frac{4E_eE_{\mathrm{ph}}}{(m_0c^2)^2}=\frac{4 \gamma E_{\mathrm{ph}}}{m_0 c^2} = 4 \gamma^2 \frac{E_{\mathrm{ph}}}{E_e}, 
\end{equation}
with $m_0$ the electron rest mass ans $E_e=\gamma m_0 c^2$.

We can distinguish the following different interaction regimes: Direct, Inverse and Symmetric Compton Scattering.

\subsection{Direct Compton}
The collision between high energy photons and electrons at rest ($E_{\mathrm{ph}} \gg T_e$, where $T_e = \left( \gamma -1 \right) m_0 c^2$) is usually called Direct Compton (DC) scattering. In this process, the photon loses energy, being red-shifted, while the collided electron gains energy, being recoiled.

The interaction studied in Arthur Compton's original experiment exploited X-rays incident on fixed target ($\beta=0$, $\gamma=1$). eq. \ref{eq:GeneralOmega} reduces therefore to:
\begin{equation}
\label{eq:DCOmega}
E_{\mathrm{ph}}^{\prime}(\theta)=\frac{E_{\mathrm{ph}}}{1 + \frac{X_{DC}}{4}(1+\cos \theta)}.
\end{equation}
In such a case, the electron recoil factor can be rewritten as a function of the well-known electron Compton wavelength  $\lambda_C=h/(m_0 c)$ and the colliding photon wavelength $\lambda$: 
\begin{equation}
X_{DC}=\frac{4E_{\mathrm{ph}}}{m_0 c^2}=\frac{4\lambda_C}{\lambda}.
\end{equation}
where $\lambda=hc/E_{\mathrm{ph}}$ ,
 leading directly to Compton's relationship for the scattered photon wavelength $\lambda^{\prime}$:

\begin{equation}
\lambda^{\prime}-\lambda=\lambda_C (1+\cos{\theta}).
\end{equation}

Here we start seeing a clear signature of radiation emission in collisions: the angular dependence of the scattered photon energy. Also, this is the fundamental expression for the photon wavelength red-shift that allowed Arthur Compton to invoke the quantum nature of the photon-electron collision in order to explain the experimental data.

\subsection{Inverse Compton Scattering}
The Inverse Compton Scattering (ICS) regime is instead characterized by collisions of highly energetic electrons and low energy photons (usually delivered by a laser) with $E_{\mathrm{ph}}\ll T_e$. 
In the interaction the photon receives energy from the electron.
ICS sources are characterized by high electron beam energy (from tens to hundreds MeV) i.e. $\beta\rightarrow 1$ and $\gamma \gg 1$.
In this conditions, and in low recoil regime, the angular aperture of the cone containing half of the emitted radiation scales as $\gamma^{-1}$. The boost effect contracts the angular coordinate of the photons, compressing such a cone.
For $\theta$ small, $cos(\theta)\approx 1 -\theta^2/2$ and eq. \ref{eq:GeneralOmega} can be approximated with:
\begin{equation}
E_{\mathrm{ph}}^{\prime}(\theta)=\frac{4E_{\mathrm{ph}}\gamma^2}{1 + X + \gamma^2\theta^2}.
\end{equation}

Equations 1 and 6 show the intrinsic dependence of the scattered photon energy on the scattering angle through the term $\theta\gamma$ in the denominator.

In the DC regime, $\gamma\simeq 1$ and the angular dependence appears without any boost effect. 

The maximum photon energy is achieved by fully back-scattered photons at $\theta=0$: $E_{\mathrm{ph}}^{\prime}(\theta=0)={4E_{\mathrm{ph}}\gamma^2}/(1 + X).$ 
In case of negligible recoil, that is in Thomson regime, the maximum photon energy just reduces to  $E_{\mathrm{ph}}^{\prime}(\theta=0)=4E_{\mathrm{ph}}\gamma^2$.

\subsection{Symmetric Compton Scattering}
In the transition between the two regimes previously discussed, the energy/momentum transfer between photons and electrons is balanced.
We call Symmetric Compton Scattering (SCS) this regime of transition between DC and ICS.
SCS is uniquely characterized by the disappearance of the energy-angle correlation of the scattered radiation that becomes monochromatic. Regarding the angular distribution, the radiation fills the whole solid angle with a different probability set by the angular dependence of the cross-section. 
Referring to equation 1, the condition for eliminating the $\theta$ dependence in $E_{\mathrm{ph}}^{\prime}$ is
\begin{equation}
\frac{X}{4} = \beta\gamma^2,
\end{equation}
a condition achieved when photon and electron energies satisfy the relation $E_{\mathrm{ph}} = \beta E_e$, which is equivalent to a photon and an electron with equal momenta and opposite directions $\vv{p}_e=-\vv{p}_\mathrm{ph}$.

Moreover, we introduce here an asymmetry factor 
\begin{equation}
    \label{Eq:A}
    A = \beta\gamma^2 - \frac{X}{4}
\end{equation}
that vanishes ($A=0$) in SCS regime and assumes large positive values ($A\rightarrow \gamma^2$) in ICS regime. 
It is worth noting that the condition $A=0$ actually deletes the angular dependence shown in eq. 1. 

In the SCS regime, the center of mass of the collisions is at rest in the laboratory reference frame (see next chapter), therefore the produced radiation is not Lorentz transformed and its frequency is no more boosted.
As a result, the energy of the scattered photons is: 
\begin{equation}
    E^\prime_{\mathrm{ph}} = E_{\mathrm{ph}}\  \forall \theta.
\end{equation}

For convenience we call $E_{0}^{\prime}$ the value that eq. \ref{eq:GeneralOmega} assumes for $\theta=0$: 
\begin{equation}
\label{eq:Taylor}
E_{0}^{\prime}=E_{\mathrm{ph}}^{\prime}(\theta=0)= \frac{2\gamma^2 \left(1+\beta \right)}{2\gamma^2 \left(1-\beta \right) + X}E_{\mathrm{ph}}.
\end{equation}
Expressing eq.  \ref{eq:GeneralOmega} in Taylor Expansion of $\theta$ around $\theta=0$ we find the following first two terms:
\begin{equation}
\label{eq:Taylor}
E_{\mathrm{ph}}^{\prime}(\theta) = E_{0}^{\prime} - \frac{E_{0}^{\prime}}{\gamma^2} \frac{A \gamma^2 \theta^2}{2\gamma^2 \left( 1-\beta \right) + X} + A\ O(\gamma^4\theta^4),
\end{equation}
where the higher order terms confirm that the symmetry condition ($A=0$) cancels the angular correlation.  

\begin{figure}[!ht]
    \centering
    \includegraphics*[width=0.85\columnwidth]{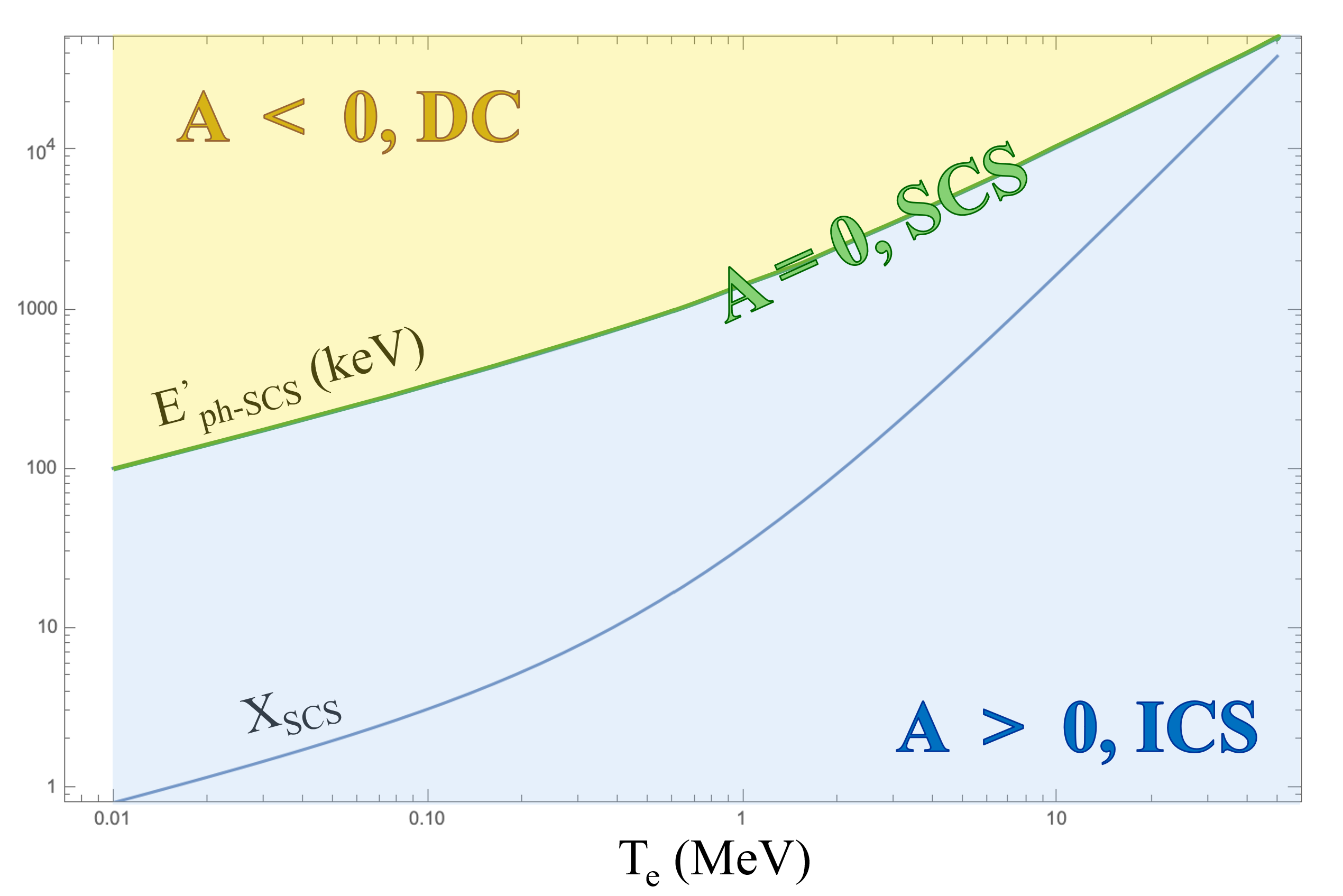}
    \caption{Evolution of the recoil factor X value and of the scattered photon energy in SCS regime as a function of the electron kinetic energy $T_e$.
    Colored areas identify the possible Compton Scattering regimes and the relative asymmetry factor A sign, DC in yellow ($A<0$), ICS in blue ($A>0$) and in green the SCS divide line ($A=0$).}
    \label{fig:Areas}
\end{figure}

\begin{figure}[!ht]
    \centering
    \includegraphics*[width=0.85\columnwidth]{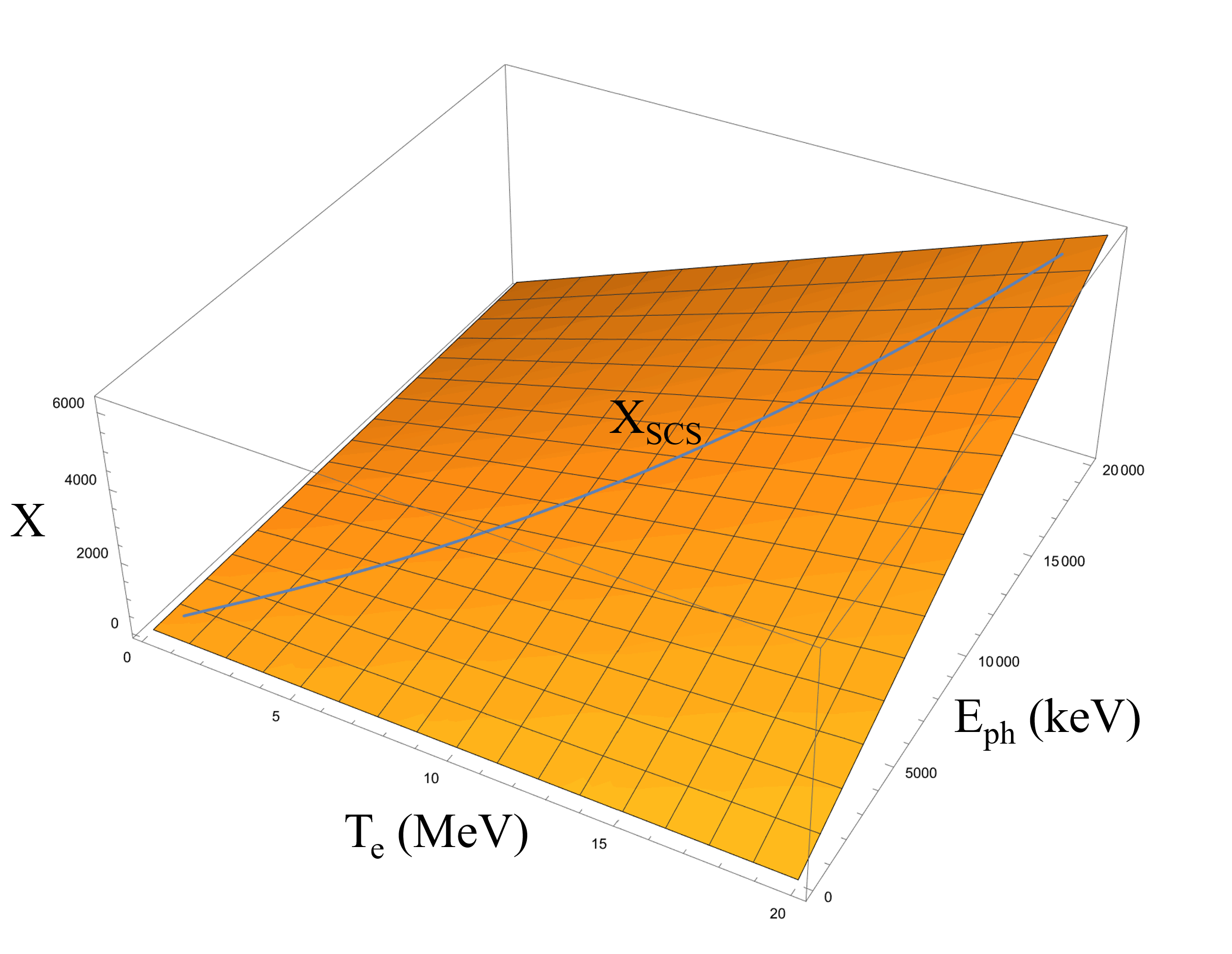}
    \caption{3D representation of the value of the recoil factor X as a function of the interacting electron kinetic energy ($T_e$) and of the incident photon energy. The line shows the recoil value in SCS conditions}
    \label{fig:X}
\end{figure}

Note that the asymmetry factor A is negative in DC regime, where $\beta=0$ and $\gamma=1$, and $A = -\lambda_C / \lambda$. On the other side, in ICS regime the asymmetry factor A is positive and scales like $\gamma^2$. 
Equation \ref{eq:Taylor} is actually a generalization of a well known formula in ICS, that reads $d E^\prime_{\mathrm{ph}}/E^\prime_0 = \gamma^2 \theta^2 /(1+X)$.  
Figures \ref{fig:Areas} and \ref{fig:X} show the dependence of $E^\prime_0$ vs. $T_e$ and of the recoil factor $X$ in different regimes (DC, SCS, ICS), that are associated to the sign of the asymmetry factor A.

\section{A Four-vector description}
The four-momentum of a particle is defined as $\textbf{p} = \left(\frac{E}{c}, p_x, p_y, p_z\right)$, where $E$ is the total energy of the particle, $c$ is the speed of light in vacuum, and $p_x, p_y, p_z$ are the components of the particle's momentum along the $x$, $y$, $z$ axes respectively.

Let us consider the case of an head-on collision between a photon and a counter-propagating electron along the z-axis.
Before the collision, the electron and the photon have the following four-momenta: 

\begin{equation}
\begin{array}{@{}r@{\;}c@{\;}l@{}}
\mathbf{p_e} &=& \left(\gamma m_0 c,0, 0, \beta \gamma m_0c \right), \\ 
\mathbf{p_{\mathrm{ph}}} &=& \left(\frac{E_{\mathrm{ph}}}{c},0, 0, -\frac{E_{\mathrm{ph}}}{c} \right).
\end{array}
\end{equation}
and the total four-momentum is:
\begin{equation}
\mathbf{p_{tot}} = \left(\gamma m_0c + \frac{E_{\mathrm{ph}}}{c},0,0,\beta\gamma m_0c - \frac{E_{\mathrm{ph}}}{c}\right).
\end{equation}

The energy available in the center of mass $E_{cm}$, in terms of the recoil factor introduced in eq. \ref{eq:GeneralRecoil}, is:
\begin{equation}
\label{eq:Ecm}
    E_{cm} = c \sqrt{\mathbf{p_{tot}} \cdot \mathbf{p_{tot}}} = m_0c^2 \sqrt{(1+\beta)\frac{X}{2}+1}=m_0 c^2 \sqrt{(1+\beta)\frac{2E_eE_{\mathrm{ph}}}{(m_0c^2)^2}+1}.
\end{equation}
The different regimes of Compton scattering can be analyzed in terms of their center of mass energy $E_{cm}$.

For the  DC regime ($\beta=0$, $\gamma=1$):

\begin{equation}
E_{cm-DC}= m_0c^2 \sqrt{\frac{2E_{\mathrm{ph}}}{m_0c^2}+1}.
\end{equation}

On the opposite side, in the ICS regime  ($\beta\simeq 1$), we obtain:
\begin{equation}
    E_{cm-ICS} = m_0c^2 \sqrt{X+1} = m_0c^2 \sqrt{\frac{4 \gamma E_{\mathrm{ph}}}{m_0c^2}+1}.
\end{equation}

Finally, for the SCS regime ($E_{\mathrm{ph}} = \beta E_e = \beta \gamma m_0c^2$):
\begin{equation}
E_{cm-SCS} = ( 1 + \beta ) \gamma m_0c^2.
\end{equation}
In this peculiar situation $E_{cm}\propto \gamma$ like in a collider. 
Being $\gamma_{cm}\equiv E_{lab}/E_{cm}$ the Lorentz boost factor associated to the center of mass reference frame, in SCS we have $\gamma_{cm}=1$ (because $E_{lab-SCS} = E_{cm-SCS}$), meaning that the center of mass of the system is at rest in the laboratory system, and the radiation produced here has the same angular and spectral distribution seen by a detector at rest in the lab.
On the other hand, DC and ICS exhibit a dependence of the available energy $E_{cm}$ typical of a fixed target collision, where $E_{cm}$ scales like $E_{cm} \propto \sqrt{T_p}$, where $T_p$ is the projectile kinetic energy. 
ICS regime is characterized by $\gamma_{cm} \gg 1$ since the center of mass reference frame is almost traveling with the electron (as shown in ref. \cite{Recoil17} $\gamma_{cm}= \gamma/(1+X)$).

\section{Effects of deep recoil in Compton scattering}
The energy spread of the scattered photon beam ($d E^\prime_{\mathrm{ph}}/E^\prime_{\mathrm{ph}}$, that is typically referred as relative bandwidth) has a vanishing dependence on the energy spread of incident photon beam ($d E_{\mathrm{ph}}/E_{\mathrm{ph}}$) whenever the recoil factor is very large.
This effect is clearly illustrated in ref \cite{Compton18}, eq. 9 for the Compton scattering interaction:
\begin{equation}
\label{Eq:DeepRanjan}
\frac{dE^\prime_{ph}}{E^\prime_{ph}} = \frac{2+X}{1+X}\frac{d\gamma}{\gamma}+\frac{1}{1+X}\frac{dE_{ph}}{E_{ph}}
\end{equation}
In this equation, the impact of high recoil factor values of X can be seen in the form of damping of the dependence of energy spread for the scattered photon beam ($\frac{dE^\prime_{ph}}{E^\prime_{ph}}$) on the energy spread of the incident photon beam ($\frac{dE_{ph}}{E_{ph}}$) and a resulting that's equal to the energy spread of the incident electrons ($\frac{d\gamma}{\gamma}$).
\begin{figure}[!ht]
    \centering
    \includegraphics*[width=.9\columnwidth]{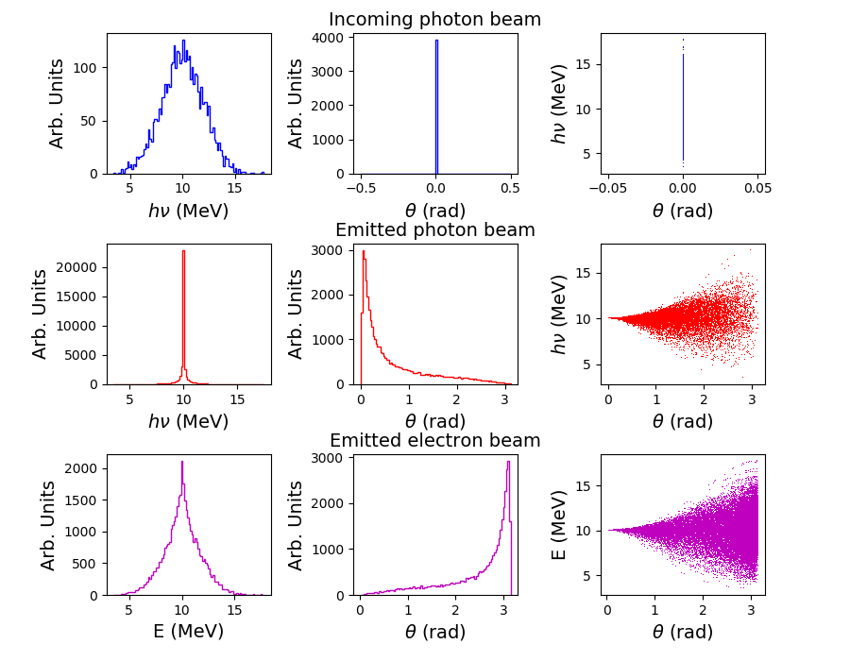}
    \caption{Simulations of SCS are shown through 9 plots arranged in three rows representing the energy and angular distributions of the particles involved in the interaction. The rows depict the incident photons, outgoing photons, and outgoing electrons, respectively. In the first column, the energy distributions of the three particle species are displayed. In the second column, the angular distributions of the outgoing particles are shown. In the third column, the energy of the particles are shown as a function of their propagation angle.}
    \label{fig:Whizard}
\end{figure}

We derive the dependence of the outgoing photon energy spread on the incident photon energy spread to study the effect of deep recoil:
\begin{equation}
\label{Eq:Band-Band}
    \frac{d E_0^\prime}{E_0^\prime} = \frac{\frac{2}{1+\beta}}{2(1-\beta)\gamma^2 + X} \frac{d E_0}{E_0}= \frac{1}{1 + \frac{1+\beta}{2} X} \frac{d E_0}{E_0}
\end{equation}
for $\beta\rightarrow1$ this result reproduces the second term of the right hand side of eq. \ref{Eq:DeepRanjan}.

We also derive the dependence of the outgoing photon energy spread on the energy spread of the incoming electron beam under the approximation of $\gamma \gg 1$, $\beta \simeq 1 - 1/2\gamma^2$
\begin{equation}
\label{Eq:Band-Esp}
    \frac{d E_0^\prime}{E_0^\prime} = \frac{1}{1+\frac{1}{4\gamma^2}} \frac{2X + X^2 + X/\gamma}{X \left( 1+X \right)} \frac{d \gamma}{\gamma}
\end{equation}
This result reproduces the first term of the right hand side of eq. \ref{Eq:DeepRanjan} at the limit $\gamma\gg 1$ .

%We also checked that the characteristics of the photon beam is fairly insensitive to the divergence of the incoming photon beam due to its transverse emittance.

%We generalize the expression given in eq. 10 of reference \cite{Compton18} by calculating the relative increment of eq. \ref{eq:GeneralOmega} around $\theta = 0$ with the second derivative respect the angle $\theta$ :
%\begin{equation} 
%\left.\frac{\Delta E^{\prime}_{ph}}{E^{\prime}_{ph}}\right|_{\theta\rightarrow0} = - \frac{\left(\beta \gamma^2 - \frac{X}{4}\right) \theta^2}{2\left(1-\beta\right)\gamma^2+X}\frac{d\theta}{\theta} = -\frac{A\  \theta^2}{\frac{2}{1+\beta} + X} \frac{d\theta}{\theta}.
%\end{equation}
%This equation is clearly zero in symmetric scattering regime where $A=0$ and simplifies to
%\begin{equation} 
%\label{eq:angularSCS}
%\left.\frac{\Delta E^\prime_{ph}}{E^\prime_{ph}}\right|_{\theta\rightarrow0} = \frac{\left(\gamma\theta\right)^2}{1+X}\frac{d\theta}{\theta}
%\end{equation}
%for $\beta\rightarrow1$ consistent with eq. 10 of reference \cite{Compton18} when the scattering is not strictly symmetric.
%The important consequence of eq. \ref{eq:angularSCS} is that for nonzero incidence angle of the photons the dependence on that angle is damped by the recoil factor X. (NON NE SONO SICURO!) 

\section{Symmetric Compton Scattering simulation}
This chapter is focused on the simulations of the Symmetric Compton Scattering using two different codes to quantify the phenomenon.
Our simulation approach investigates energy transfer, scattering angles and bandwidth variation of the particles that interacted.

\subsection{WHIZARD}
We used the WHIZARD code \cite{Kilian2011}, a universal parton-level Monte Carlo event generator, to perform simulations of SCS.

An almost monochromatic (with an rms energy spread of the order of $10^{-4}$) 10 MeV electron beam ($\beta\rightarrow 1$) collided head-on with an incoming photon beam characterized by large bandwidth (20 percent rms spread). 
The recoil in this interaction is large, the computed value being $X=1533$.
The results of the interaction, shown in fig. \ref{fig:Whizard}, confirmed the theoretical predictions: the outgoing photons showed no correlation between energy and emission angle and featured a significant narrowing of the bandwidth ($2\cdot10^{-4}$ rms spread, i.e. a reduction of the energy spread by about 3 orders of magnitude from incident photon beam to the scattered photon beam). 
On the other hand, the electron beam emerging from the interaction inherited an high energy spread (of the order of $10^{-1}$) from the original interacting photon beam, displaying an entropy exchange.

\begin{figure}[!ht]
    \centering
    \includegraphics*[width=.9\columnwidth]{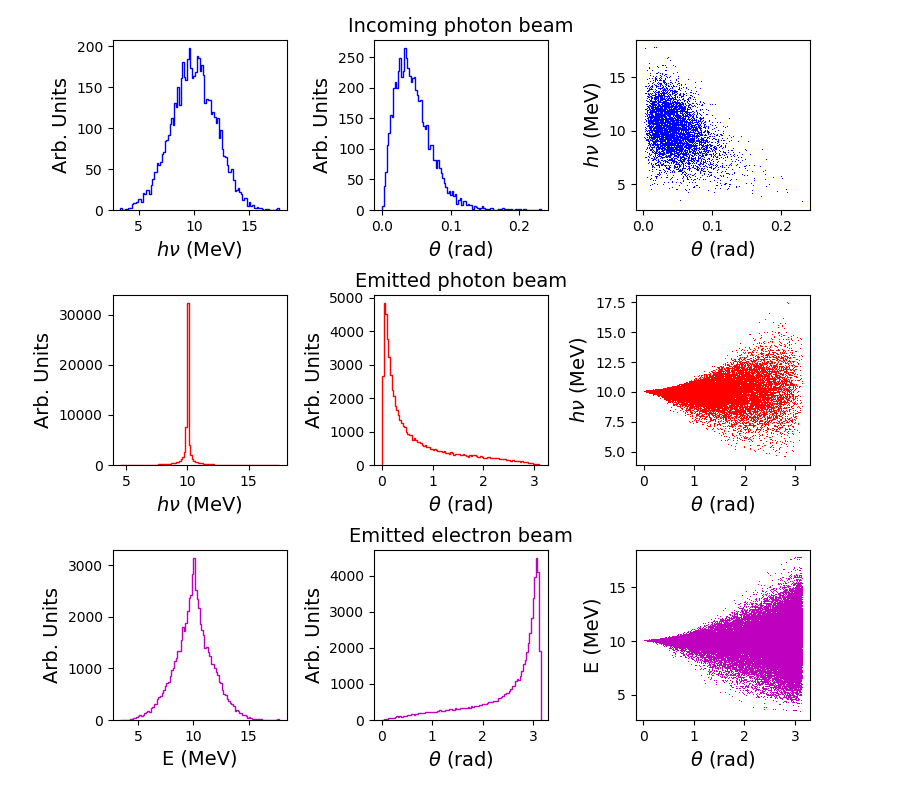}
    \caption{Simulations of SCS with an incoming photon beam displaying a correlation between angle of propagation and photon energy. The results are shown through 9 plots arranged in three rows as in fig. \ref{fig:Whizard}. The angular correlation of the incoming photon beam is removed in the interaction thanks to the high recoil factor ($X \sim 1500$).}
    \label{fig:WhizardAng}
\end{figure}

WHIZARD was also used to perform an analysis of the SCS effect in presence of angular divergence of the incident photon beam, shown in fig. \ref{fig:WhizardAng}, by mixing several runs with different scattering angle. The result confirms the SCS  mono-chromatization also in interactions characterized by small incidence angles.

\subsection{Monte Carlo code}
\begin{figure}[!ht]
    \centering
    \includegraphics*[width=\columnwidth]{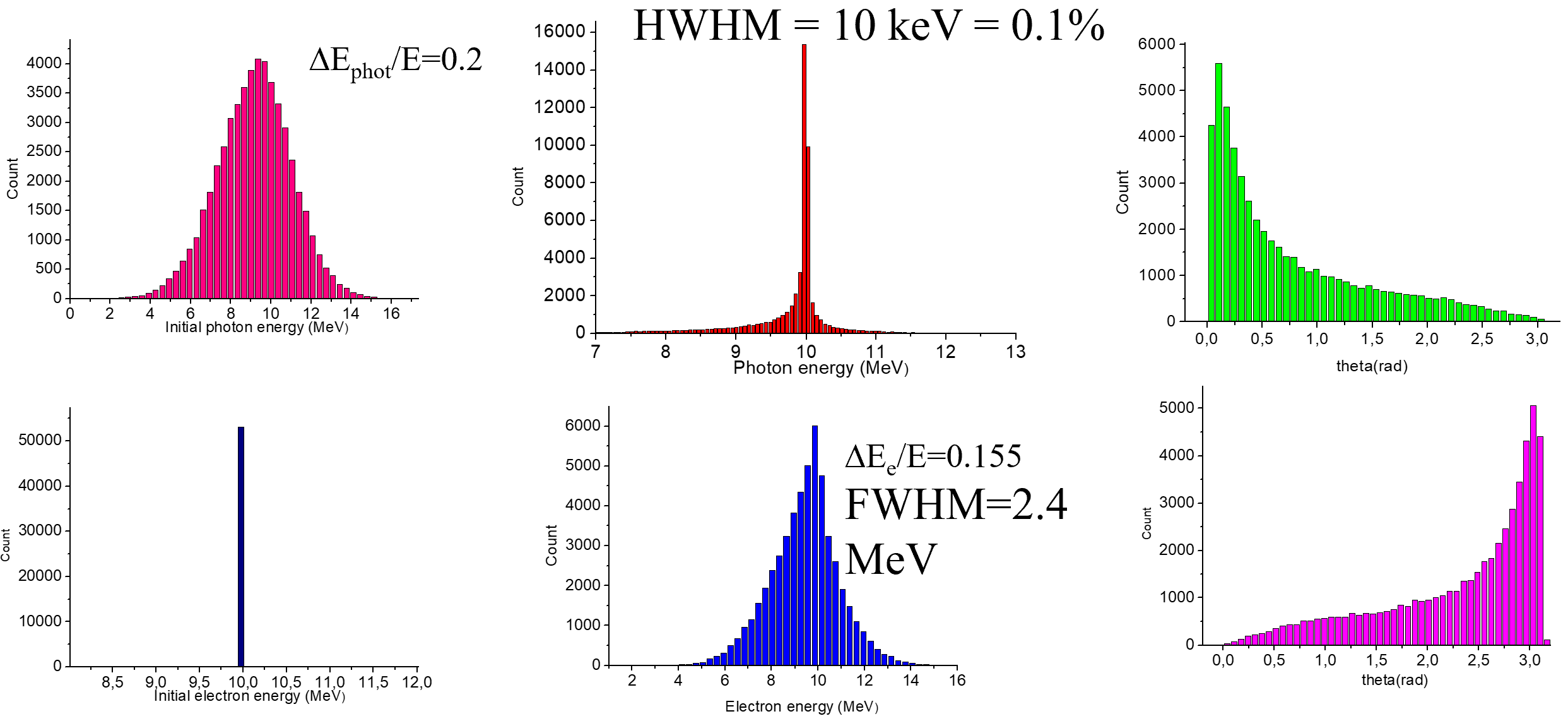}
    \caption{This figure illustrates the simulations of the SCS regime for a recoil factor of $X=1533$. The plot consists of six subplots organized into two rows, with the first row showing the photon behavior and the second row showing the electron behavior.
    In each row, the leftmost column displays the bandwidth of the incoming bunch, the central column shows the resulting particles after the interaction, and the rightmost column displays the angular distribution of the particles. This arrangement allows for a detailed analysis of the behavior of photons and electrons during the SCS regime.}
    \label{fig:SCSVittoria}
\end{figure}

A home made multitasking Monte Carlo code has been developed, validated for different type of collisions  and applied to the Compton scattering process \cite{tesi}. 
As an additional internal feature, the code allows to consider the energy and angular (polar and azimuth) spread of both incident beams.
To confirm the occurrence of the effect, we conducted the same simulation of the deep-recoil SCS interaction (at $X=1533$). 
Our findings confirm the exchange of entropy, resulting in a reduction of the bandwidth of the emitted radiation and an enlargement of the electron's bandwidth (as summarized in fig. \ref{fig:SCSVittoria}).

\begin{figure}[!ht]
    \centering
    \includegraphics*[width=\columnwidth]{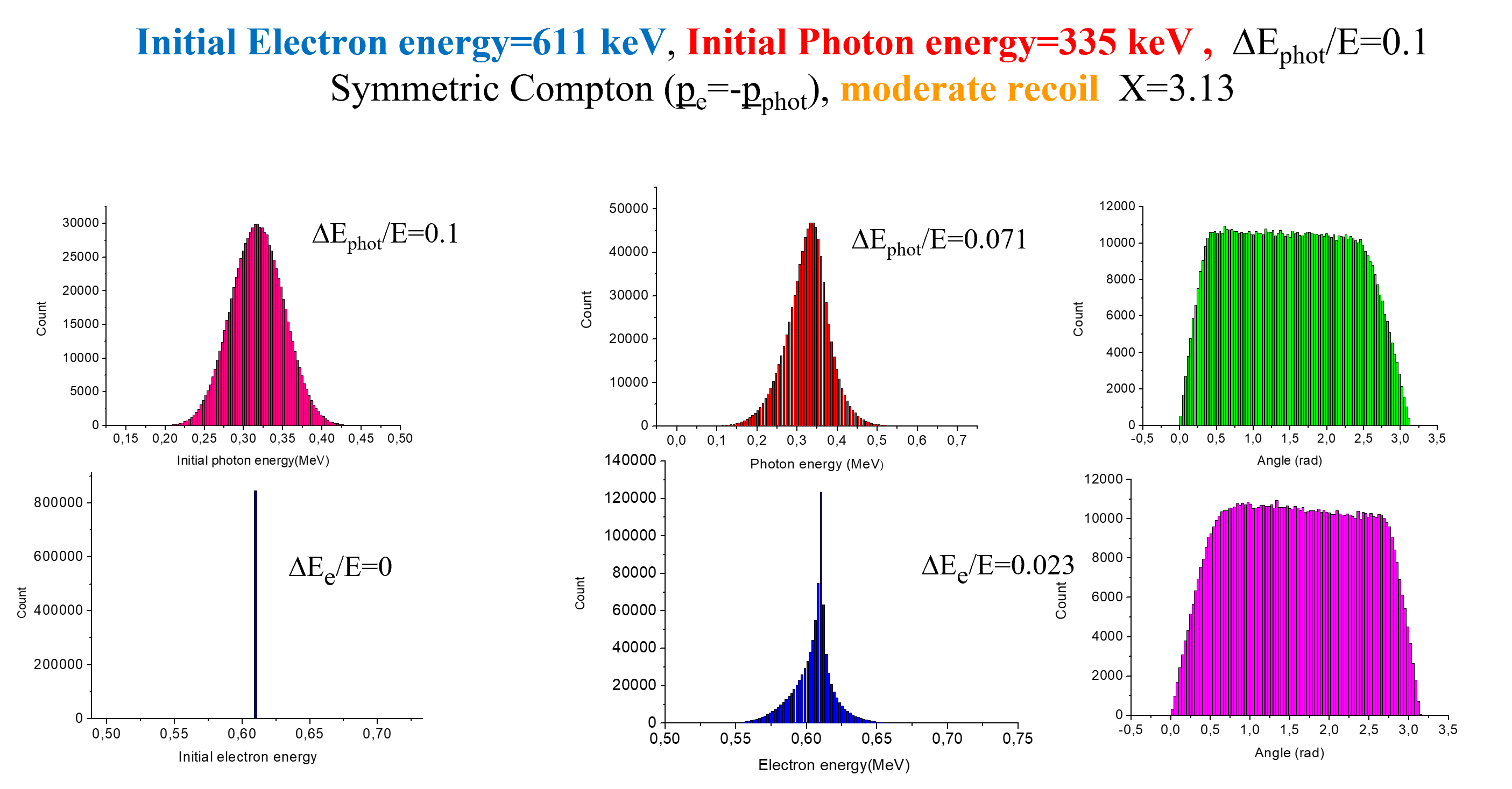}
    \caption{This figure shows the simulations of the SCS regime for a small recoil factor of $X\sim 3$. The plot consists of six subplots organized into two rows, with the first row showing the photon behavior and the second row showing the electron behavior.
    In each row, the leftmost column displays the bandwidth of the incoming bunch, the central column shows the resulting particles after the interaction, and the rightmost column displays the angular distribution of the particles. This arrangement allows to underline the incomplete exchange of energy spread due to the reduced recoil factor.}
    \label{fig:Trans}
\end{figure}

Furthermore, we examined the transition from the SCS regime to the ICS regime, with particular focus on the angular distribution of the scattered radiation. 
To explore the transition regime, we started with the deep-recoil SCS interaction ($X=1533$) and slightly increased the energy of the incident electron bunch while reducing the energy of the photon bunch. 
We investigated three cases, specifically with electron-photon energies of ($E_e \simeq E_{ph} = 10$ MeV), ($E_e = 11$ MeV, $E_{ph} = 9.08$ MeV), and ($E_e = 12$ MeV, $E_{ph} = 8.33$ MeV).
The results, depicted in fig. \ref{fig:Trans}, show the distribution shifting from an uncorrelated energy-angle pattern to a more correlated one, resembling the typical "mustache" shaped curve observed in ICS experiments.
\begin{figure}[!ht]
    \centering
    \includegraphics*[width=\columnwidth]{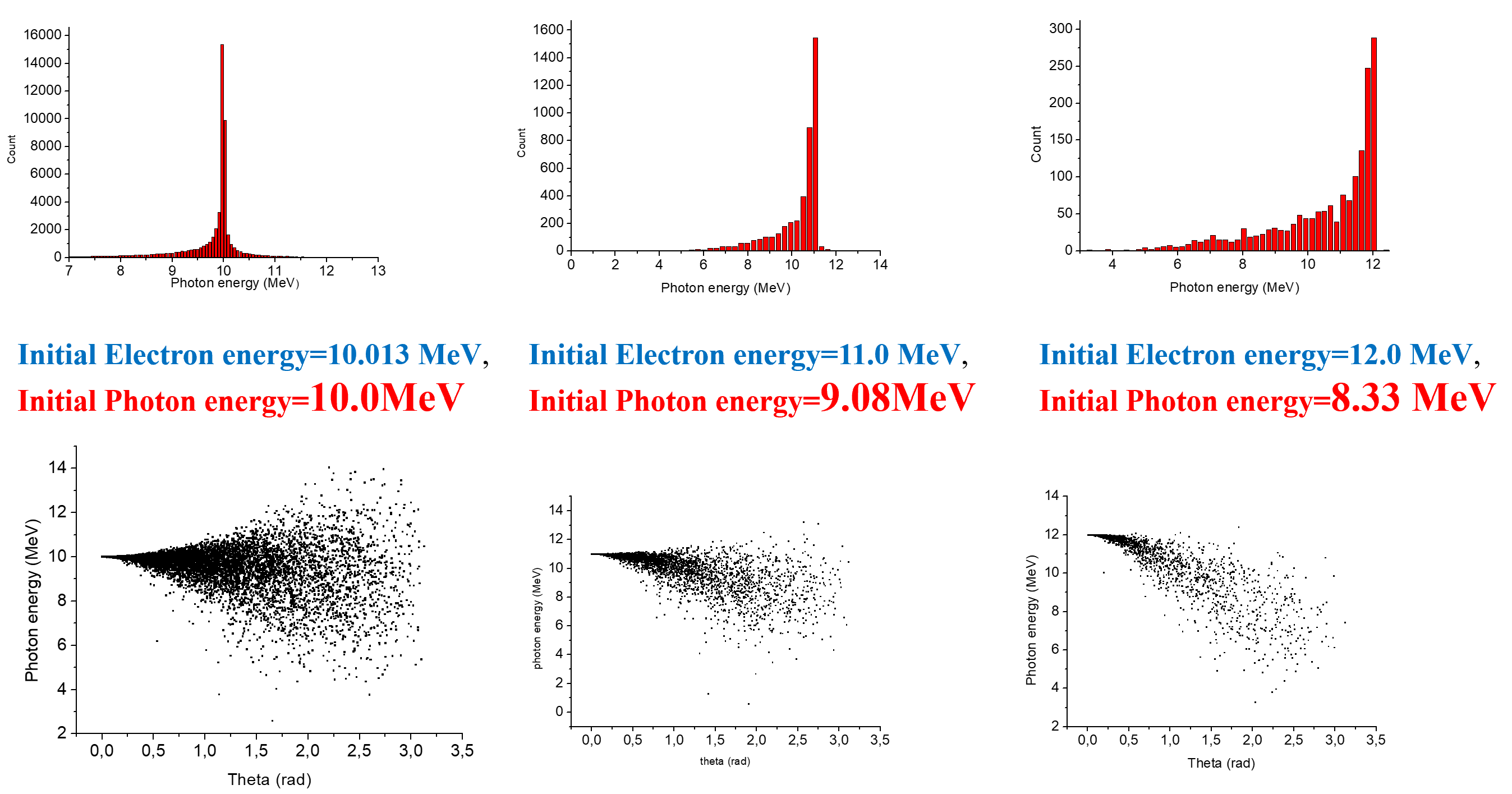}
    \caption{This figure demonstrates the regime transitions between SCS and ICS for three different sets of photons and electrons energy.
    The upper row displays the produced photon energy distribution for each set, while the bottom row shows the angular distribution of the particles (i.e., energy as a function of emission angle) for each set. This arrangement allows for a comprehensive comparison of the behavior of the particles during the transition between SCS and ICS regimes.}
    \label{fig:Mild}
\end{figure}

Finally we investigated the SCS regime with low recoil factor ($A=0$, $X=3.13$). 
In this peculiar situation (see fig. \ref{fig:Mild}), a milder mono-chromatization  of the incident photons occurs, compared with what happens in deep recoil regime represented in. \ref{fig:SCSVittoria}. 
\section{Conclusions}
We explore the transition between Compton Scattering and Inverse Compton Scattering (ICS), a regime characterized by an equal exchange of energy and momentum between the colliding particles. This regime has been called Symmetric Compton Scattering (SCS) and has the unique property of transferring mono-chromaticity from one beam to the other, resulting in back-scattered photons that are intrinsically monochromatic. The paper suggests that large-recoil SCS or quasi-SCS can be used to design compact intrinsic monochromatic gamma-ray sources, thus avoiding the use of GeV-class electron beams and powerful laser/optical systems typically required for ICS sources. 

Indeed the capability of SCS regime to vanish the photon energy-angle correlation, married to the large recoil beneficial effects on the scattered photon energy spread, as shown in previous chapters, makes possible to conceive a mono-chromatic gamma ray beam source based on the collision between a bremsstrahlung radiation beam (or a coherent bremsstrahlung beam from a channeling source, \cite{Dabagov2008}) and a mono-energetic electron beam of similar energy, say in the 2-10 MeV range, so to employ a compact source that is much more sustainable than typical ICS sources for nuclear physics/photonics like ELI-NP-GBS \cite{Eli}, that all envisage the use of GeV-class linear accelerators.

An Energy Recovery Linac with 10-20 MeV electron beam energy would allow to sustain a much larger average current (in the range of tens of mA) than a room temperature Linac like in ELI-NP-GBS, to the collision point with the broad-band bremsstrahlung photon beam, compensating the decrease of total cross section $\sigma$ with the recoil factor $X$ typical of Klein-Nishina formula, as shown here below (see ref. \cite{Recoil17})
\begin{equation}
\label{eq:Cross}
  \left\{
    \begin{aligned}
      \lim\limits_{X \to 0} \sigma &= \frac{8 \pi r_e^2}{3}(1-X) = \sigma_T(1-X) \\
      \lim\limits_{X \to \infty} \sigma &= \frac{2 \pi r_e^2}{X}\left(\log X+\frac{1}{2}\right)
    \end{aligned}
  \right.
\end{equation}
A detailed study of a SCS $\gamma$-ray source will be the subject of a future work, that will have to take into account the compensation of the cross section decrease for large recoils with the positive effects of reducing the photon bandwidth by large recoils, as shown by equations \ref{Eq:DeepRanjan}, \ref{Eq:Band-Band} and \ref{Eq:Band-Esp} (assuming the capability to bring to collision a good quality electron beam with small energy spread $\frac{\Delta_\gamma}{\gamma}$ below $10^{-3}$ \cite{JAP}). 
The guidelines of such a design study should be oriented to optimize the SCS $\gamma$-ray source in terms of maximum Spectral Density $S_d$, as illustrated in \cite{JAP} and typically requested by nuclear photonics and photo-nuclear physics applications. 
$S_d$ is actually defined as $S_d \equiv N_\mathrm{ph} (\Delta E^\prime_{ph}/E^\prime_{ph})^{-1}$, where $N_\mathrm{ph}$ is the number of gamma ray photons generated per second within the relative bandwidth $\Delta E^\prime_{ph}/E^\prime_{ph}$ around the nominal average energy $E^\prime_{ph}$. 
Since $N_\mathrm{ph}$ scales with the product $L \sigma$, where $L$ is the collision luminosity and $\sigma$ is given in eq. \ref{eq:Cross}, we see that for large recoil collisions $S_d$ scales like $L \log{X}/X$. 
On the other side $\Delta E^\prime_{ph}$ becomes very small in case of SCS at large recoil, as stated by equations \ref{eq:Taylor}, \ref{Eq:Band-Band} and \ref{Eq:Band-Esp}, and well illustrated in figures \ref{fig:Whizard}, \ref{fig:WhizardAng} and \ref{fig:SCSVittoria}, showing the potentiality of SCS to attain photon beams with relative bandwidths smaller than $10^{-3}$.

\bibliographystyle{unsrt}
%

%\bibliography{references}  %%% Uncomment this line and comment out the ``thebibliography'' section below to use 

\end{document}